\documentclass[preprint,showpacs,preprintnumbers,amsmath,amssymb]{revtex4}

\usepackage{graphicx}
\usepackage{dcolumn}
\usepackage{bm}

\begin{document}


\title{Quantum Vacuum Contribution to the Momentum of the Dielectric Media.}
\author{A. Feigel}
\affiliation{Electro-Optics Division, Soreq NRC, Yavne 81800, Israel}

\begin{abstract}
Momentum transfer between matter and electromagnetic field is analyzed. The
related equations of motion and conservation laws are derived using
relativistic formalism. Their correspondence to various, at first sight
self-contradicting, experimental data (the so called Abraham-Minkowski
controversy) is demonstrated. A new, Casimir like, quantum phenomenon is
predicted: contribution of vacuum fluctuations to the motion of dielectric
liquids in crossed electric and magnetic fields. Velocities about $50nm/s$ can
be expected due to the contribution of high frequency vacuum modes.

\end{abstract}
\maketitle

Electromagnetic radiation possesses energy, linear and angular momenta like
any ordinary material object. As a consequence the light can exert mechanical
forces on matter during the interaction. The forces associated with exchange
of the linear and the angular momenta of light with material bodies were
successfully measured already hundred years ago\cite{leb}\cite{nich}%
\cite{beth}. However, the fundamental question of the momentum associated with
a photon in an optically dense medium is still under discussion\cite{loudon},
despite that it was formulated in the beginning of the previous century. This
problem arises from the discrepancy between Minkowski's\cite{minko}
$G_{M}=\frac{1}{4\pi c}\int d^{3}xD\times B$ and Abraham's\cite{abrah}
$G_{A}=\frac{1}{4\pi c}\int d^{3}xE\times H$ results, where $E$, $H$, $D$ and
$B$ are electric and magnetic fields and inductances correspondingly. Their
difference is significant: while Minkowski's moment is directly proportional
to the refractive index of the medium, the moment of Abraham possess inverse proportionality.

Minkowski momentum is considered by many as unacceptable, although it was
shown by Jones, Richards and Leslie (JRL) that the recoil force of the light
on a mirror immersed in liquid is directly proportional to the refractive
index of the liquid\cite{jones1}. This experiment was conducted twice with a
20 years period\cite{jones2}. However most of the theoretical works are in
favor of Abraham's expression (for a review, see refs. \cite{pierls}%
\cite{brevik}).

Abraham's momentum can be derived from the Poynting energy flow vector
$S=\frac{c}{4\pi}E\times H$ under assumption that all energy is purely
electromagnetic and relates to the mass through the $U=mc^{2}$ relativistic
formula. It corresponds to the relativistic requirement for direct
proportionality of the energy and momentum flows (the symmetry of the
electromagnetic stress tensor)\cite{jackson}. Abraham's result is also
supported by statistical physics approaches\cite{groot}. However, as far as we
know, there are no experimental data that demonstrate the inverse dependence
of the radiation pressure on the refractive index. This, at least, allows to
conclude that the measured momentum is not purely electromagnetic.

The effective momentum of a photon in a dielectric medium consists of
electromagnetic momentum and associated motion or even radiation of
matter\cite{pierls}. However separate identification of different parts proved
itself to be non-trivial and sometimes led to contradictions with experimental
data\cite{jones2}. Blount\cite{Blount} and Nelson\cite{nelson} developed a
Lagrangian formalism of the problem, using heuristic and microscopic averaging
approaches correspondingly. They significantly clarified the picture by
associating Abraham's expression with electromagnetic momentum and Minkowski's
momentum with phonon like pseudomomentum. Still several questions remained
open, especially a small discrepancy of Abraham's momentum with the expression
derived in refs. \cite{nelson}\cite{lodnel}.

In this Letter the related Lagrangian and corresponding equations of motion
are derived using relativistic formalism. In the case of liquid dielectric,
interaction of the electromagnetic field with matter causes motion of the
latter. Thus while Abraham's expression is indeed the momentum of the field,
the measured momentum also includes the matter contribution and its value
coincides with Minkowski's result. Afterwards the possible vacuum
contributions to the motion of the matter are considered. Each electromagnetic
mode possesses finite momentum, even in its ground state\cite{mandel}. Thus
modification of the modes by matter can alter the momentum of the vacuum. The
latter generally vanishes due to counter propagating modes that cancel each
other's contribution. This situation can be different however in materials
that are temporally and spatially asymmetric.

The electromagnetic field in an optically dense medium is described by the
Maxwell equations:%
\begin{equation}
\nabla\times H=\frac{1}{c}\frac{\partial D}{\partial t},\nabla D=0,\nabla
\times E=-\frac{1}{c}\frac{\partial B}{\partial t},\nabla B=0. \label{gch}%
\end{equation}
The free electric $E$ and magnetic $B$ fields exist both inside and outside
the matter. The matter response to the radiation is taken into account through
derived fields $D$ and $H$, related to $E$ and $B$ by the dispersion
relations. In the case of linear, non-dispersive dielectric medium they are
given by:%
\begin{equation}
D=\varepsilon E,H=\frac{1}{\mu}B, \label{d=e0}%
\end{equation}
where $\varepsilon$\ and $\mu$ are the dielectric and magnetic constants of
the matter correspondingly.

The Lagrangian approach simplifies the investigation of light-matter
interaction. The simplicity comes from the universality that allows to use the
same approach to the system consisting of significantly different subsystems
and the natural to the Lagrangian approach definition of the conservation
laws. The Lagrangian which is equivalent to eqs. (\ref{gch}) and (\ref{d=e0})
is\cite{lanfield}\cite{goldstein}:%
\begin{equation}
L_{Field}=\int d^{3}x\frac{1}{4\pi}\left(  \frac{\varepsilon}{2}E^{2}-\frac
{1}{2\mu}B^{2}\right)  . \label{lf=}%
\end{equation}
The first pair of the Maxwell equations (\ref{gch}) corresponds to the
equations of motion:%
\begin{equation}
\frac{\partial}{\partial t}\frac{\partial L_{Field}}{\partial\frac{\partial
A_{i}}{\partial t}}=\frac{\partial L_{Field}}{\partial A_{i}},\frac{\partial
}{\partial t}\frac{\partial L_{Field}}{\partial\frac{\partial\Phi}{\partial
t}}=\frac{\partial L_{Field}}{\partial\Phi}, \label{ddl}%
\end{equation}
while the second pair of (\ref{gch}) follows from definitions of the vector
$A$ and the scalar $\Phi$ potentials:%
\begin{equation}
E=-\frac{1}{c}\frac{\partial A}{\partial t}-\nabla\Phi,B=\nabla\times A.
\label{e=-}%
\end{equation}

The motion of the matter and especially its influence on the electromagnetic
field must be taken into account in a combined matter-field Lagrangian
$L_{MF}$. The linear dispersion relations (\ref{d=e0}) change in moving media
to:%
\begin{equation}
D=\varepsilon E+\frac{\varepsilon\mu-1}{c\mu}v\times B,B=\mu H+\frac
{\varepsilon\mu-1}{c}E\times v, \label{d=ee}%
\end{equation}
where first order $v/c$ terms were taken into account\cite{landel}%
\cite{pauli}. They follow from relativistic requirements and can be derived
using the first order Lorentz transformations:%
\begin{equation}
E\rightarrow E+\frac{1}{c}v\times B,B\rightarrow B+\frac{1}{c}E\times v,
\label{ese}%
\end{equation}%
\begin{equation}
D\rightarrow D+\frac{1}{c}v\times H,H\rightarrow H+\frac{1}{c}D\times v,
\label{dsd}%
\end{equation}
relative to (\ref{d=e0}). These transformations can be applied directly to the
Lagrangian (\ref{lf=}). Substituting (\ref{ese}) into (\ref{lf=}), keeping the
first order $v/c$ terms and adding $\rho v^{2}/2$ one obtains:%
\begin{equation}
L_{MF}=\int d^{3}x\left(  \frac{1}{2}\rho v^{2}+\frac{1}{4\pi}\left(
\frac{\varepsilon}{2}E^{2}-\frac{1}{2\mu}B^{2}+\frac{\varepsilon\mu-1}{\mu
c}B\left(  E\times v\right)  \right)  \right)  . \label{l=l}%
\end{equation}
Since the liquid is assumed to be incompressible, it is described by its
density $\rho$ and local velocity $v$ only. The equations of motion
(\ref{ddl}) of (\ref{l=l}) are identical to the Maxwell equations with
dispersion relations (\ref{d=ee}). The last term of (\ref{l=l}) can be
rewritten in an interaction $Aj$ form, where the current $j$ is given by
$\nabla\times\left(  E\times v\right)  \left(  \varepsilon\mu-1\right)  /\mu
c$. The latter, at least for the non-magnetic $\mu=1$ case, can be obtained by
microscopic averaging procedures\cite{nelson}\cite{jackson}.

The Lagrangian (\ref{l=l}) is explicitly independent of the space coordinates
$x$, due to the homogeneity of space. Thus according to Noether theorem, the
momentum\cite{goldstein}:%
\begin{equation}
G_{i}=\int d^{3}x\left(  \frac{\partial L}{\partial v_{i}}-\frac{\partial
L}{\partial\frac{\partial A_{j}}{\partial t}}\frac{\partial A_{j}}{\partial
x_{i}}\right)  \label{pi=}%
\end{equation}
is conserved. Substituting (\ref{l=l}) into (\ref{pi=}) one obtains:%
\begin{equation}
G=\int d^{3}x\left(  \rho v+\frac{1}{4\pi}\left(  \frac{\varepsilon}{c}E\times
B-\frac{\varepsilon\mu-1}{\mu c}E\times B\right)  \right)  =\int d^{3}x\left(
\rho v+\frac{1}{4\pi c}E\times H\right)  . \label{p=r}%
\end{equation}
The corresponding angular momentum $l=x\times G$ becomes:
\begin{equation}
l=\int d^{3}x\left(  x\times\rho v+\frac{1}{4\pi c}x\times E\times H\right)  .
\label{l=x}%
\end{equation}
Therefore the conserved linear (\ref{p=r}) and angular (\ref{l=x}) momenta
consist of the matter and the Abraham's field terms. The correspondence
between the conserved and the measured momenta follows from the analysis of
the Lorentz force acting on material objects\cite{jackson}.

The $\rho v$ term can be obtained from the liquid's equation of motion:%
\begin{equation}
\frac{\partial}{\partial t}\frac{\partial L}{\partial v}=\frac{\partial
L}{\partial R}, \label{ddlv}%
\end{equation}
where $R$ is the matter coordinate. Far from the boundaries, $\partial
L/\partial R$ can be neglected, leading to:%
\begin{equation}
\rho v=\frac{\varepsilon\mu-1}{4\pi\mu c}E\times B. \label{rv=e}%
\end{equation}
This expression corresponds qualitatively to the pseudomomentum of ref.
\cite{nelson}. Substituting (\ref{rv=e}) into (\ref{p=r}), one obtains
$G=D\times B/4\pi c$, which was observed in the JRL experiments.

By analogy, the measured angular momentum is $l=x\times\left(  D\times
B\right)  $. It can be separated into "spin" $D\times A$ and "orbital" $\sum
D_{j}\left(  x\times\nabla\right)  A_{j}$ parts. The latter does not vanish
only in the case of beams with specific wavefront distortions, e.g.
Laguerre-Gaussian beams\cite{gausslag}. In contrast to the linear momentum,
the spin part of the nearly plane wave is independent of dielectric properties
of the medium. This was verified experimentally for microwave radiation
\cite{angmicro}.

The dielectric constant dependent angular momentum was observed inside a
cylindrical capacitor filled with dielectric\cite{cannat}. A cylindrical
capacitor was suspended on torsional levers inside some external magnetic
field parallel to the axis of the cylinder. By applying voltage to the
capacitor's walls a radial electric field was created. These independent
fields possessed non-vanishing angular momentum, which was compensated by the
motion of the capacitor itself. The observed\cite{canada} $x\times\rho
v\propto\left(  \varepsilon-1\right)  l_{0}$, where $l_{0}$ corresponds to the
$\varepsilon=1$ case, follows from (\ref{rv=e})\cite{smol}.

Ashkin and Dziedzic observed that the liquid interface bends outwards the
liquid in both cases when light enters and leaves the liquid\cite{ashkin}.
Contrary to their measurement, the conservation law (\ref{p=r}) predicts
inward bending. Loudon recently arrived to the same conclusion by quantum
analysis of the Lorentz force\cite{loudon}. The results of ref. \cite{ashkin}
were explained by the influence of ponderomotive forces\cite{gordon}, caused
by strong focusing of the light in this experiment. These forces, also used
for optical tweezers, are much stronger than contributions from the change of
the radiation momentum on the boundary.

Radiation forces can be caused even by redistribution of the energy between
quantum vacuum and matter. Attraction of two parallel metal plates in vacuum
was predicted by Casimir\cite{casimir} and experimentally observed by
Lamoreaux\cite{lamor}. Electromagnetic field possesses finite energy even in
the ground state, similar to the quantum harmonic oscillator. The presence of
dielectric or metallic objects in space alternates the eigenstates of the
electromagnetic field. The energy of such system depends on the specific
arrangement of the objects and some rearrangement can be energetically
favorable. Between two metal plates the low frequency electromagnetic modes
are cut off by the boundary conditions. The smaller the separation the smaller
is the effective energy of the system, consequently the plates attract each
other. However, in Casimir case, no momentum is gained by the plate's center
of the mass according to symmetry considerations. Moreover, to the best of our
knowledge, the transfer of finite momentum from vacuum modes to matter was not
considered yet.

The zero fluctuations contribution to the equation of motion (\ref{rv=e}) can
be expected, since the moment of electromagnetic field, similar to its energy,
is a quadratic function of $E$ and $B$. Vacuum contribution can not occur,
neither in time even media nor in spatially symmetrical time-odd (Faraday)
materials, due to the self compensation of counterpropagating modes. Therefore
both time and spatial asymmetries are required.

The break of both spatial and time symmetries occur naturally in
magnetoelectric materials\cite{fuchs},\cite{dell}. The dispersion relations
for magnetoelectrics are:%
\begin{equation}
D=\widehat{\varepsilon}E+\widehat{\chi}H,B=\widehat{\mu}H+\widehat{\chi}^{T}E.
\label{d=e}%
\end{equation}
The same dispersion can be created artificially by applying external electric
and magnetic fields\cite{rothbi}. In this case, the dielectric properties of
the medium $\widehat{\varepsilon}$, $\widehat{\mu}$ and $\widehat{\chi}$
depend on the external fields $E_{ext}$ and $B_{ext}$. For the specific case
of perpendicular electric and magnetic fields acting on isotropic
material\cite{rothvac}\cite{figotin}:%
\begin{equation}
\widehat{\varepsilon}=\left(
\begin{array}
[c]{ccc}%
\varepsilon & 0 & 0\\
0 & \varepsilon & 0\\
0 & 0 & \varepsilon
\end{array}
\right)  \widehat{\mu}=\left(
\begin{array}
[c]{ccc}%
\mu & 0 & 0\\
0 & \mu & 0\\
0 & 0 & \mu
\end{array}
\right)  \widehat{\chi}=\left(
\begin{array}
[c]{ccc}%
0 & \chi_{xy} & 0\\
\chi_{yx} & 0 & 0\\
0 & 0 & 0
\end{array}
\right)  \label{e=(}%
\end{equation}
For light propagating along $z=E_{ext}\times B_{ext}$ direction, substituting
(\ref{d=e}) and (\ref{e=(}) into Maxwell equations (\ref{gch}) one
obtains\cite{figotin}:%
\begin{equation}
n_{\overrightarrow{k},1}=\sqrt{\varepsilon\mu}+\chi_{xy},n_{\overrightarrow
{k},2}=\sqrt{\varepsilon\mu}-\chi_{yx},n_{-\overrightarrow{k},1}%
=-\sqrt{\varepsilon\mu}+\chi_{xy},n_{-\overrightarrow{k},2}=-\sqrt
{\varepsilon\mu}-\chi_{yx}. \label{nk=}%
\end{equation}
and corresponding modes $\left(  E_{x},E_{y},B_{x},B_{y}\right)  $:%
\begin{equation}
\left(  1,0,0,\sqrt{\varepsilon\mu}\right)  ,\left(  0,1,-\sqrt{\varepsilon
\mu},0\right)  ,\left(  1,0,0,-\sqrt{\varepsilon\mu}\right)  ,\left(
0,1,\sqrt{\varepsilon\mu},0\right)  . \label{(ex}%
\end{equation}

In the case of magnetoelectrics (\ref{d=e}), the term $\frac{1}{\mu}%
B\widehat{\chi}^{T}E$ must be added to the Lagrangian (\ref{l=l}). Using
(\ref{ese}) one obtains:%
\begin{equation}
L_{ME}=L_{FM}+\int\frac{d^{3}x}{4\pi}\left(  \frac{1}{\mu}B\widehat{\chi}%
^{T}E+\frac{1}{\mu c}B\widehat{\chi}^{T}\left(  v\times B\right)  +\frac
{1}{\mu c}\left(  E\times v\right)  \widehat{\chi}^{T}E\right)  . \label{lme}%
\end{equation}
Equations of motion (\ref{ddl}) correspond to the dispersion relations
(\ref{d=e}) in moving media, while (\ref{ddlv}) becomes:%
\begin{equation}
\rho^{0}v=\frac{1}{4\pi}\left(  \frac{\varepsilon\mu-1}{\mu c}E\times
B+\frac{1}{\mu c}E\times\left(  \widehat{\chi}^{T}E\right)  -\frac{1}{\mu
c}B\times\left(  \widehat{\chi}B\right)  \right)  . \label{rv=}%
\end{equation}

The non-compensating moment of a pair of counterpropagating modes in $z$
direction is $\Delta p=\left(  \chi_{xy}-\chi_{yx}\right)  \left(
1+\varepsilon\mu\right)  /\left(  2\pi\mu c\right)  $. It is obtained by
substitution of (\ref{(ex}) into (\ref{rv=}). Taking into account all
contributing modes and $\Delta p\left(  \theta\right)  =$ $\Delta pcos\theta$,
we obtain:%
\begin{equation}
v=\frac{1}{\rho}\frac{1}{2\pi}\frac{1+\varepsilon\mu}{\mu c}\underset
{0,\infty}{\overset{\pi/2,2\pi/\omega_{cut}}{%
{\displaystyle\iint}
}}\Delta ncos\theta k^{2}E^{2}\frac{dk}{\pi^{2}}sin\theta d\theta, \label{v=2}%
\end{equation}
where $\Delta n=\left(  \chi_{xy}-\chi_{yx}\right)  $. The vacuum $E_{vac}%
^{2}=\hbar\omega/2$, thus (\ref{v=2}) becomes:%
\begin{equation}
v=\frac{1}{32\pi^{3}}\frac{1}{\rho}\Delta n\frac{1+\varepsilon\mu}{\mu}%
\frac{\hbar\omega_{cut}^{4}}{c^{4}}. \label{v=int}%
\end{equation}
This expression is significantly different from the Casimir effect, since it
is powered by the high frequency cut-off. The latter makes it more similar to
the Lamb shift phenomenon.

This effect (\ref{v=int}) can be evaluated quantitatively by the estimation of
the value of $\Delta n$ from the known experimental data. In crossed external
magnetic $B_{ext}$\ and electric $E_{ext}$ fields, $\Delta n$ is proportional
to magnetoelectric susceptibility $\beta_{\perp}$\cite{buck}:%
\begin{equation}
\Delta n\approx\left(  \frac{3}{2}\beta_{\perp}-\frac{1}{2}\beta_{||}\right)
E_{ext}B_{ext}l_{0}^{-1}\approx\beta_{\perp}E_{ext}B_{ext}l_{0}^{-1}
\label{dk=}%
\end{equation}
where $l_{0}\approx0.3nm$ is the characteristic interatomic distance. This
result follows from the spherically symmetric system's fourth order energy
terms $L=1/4\beta_{\perp}E^{2}B^{2}+1/4\left(  \beta_{||}-\beta_{\perp
}\right)  \left(  EB\right)  ^{2}$ and $D=\partial L/\partial E$ relation. The
$\chi_{xy}+\chi_{yx}\approx1e-11$ was recently observed \cite{rothbi} by
measurement of magnetoelectric linear birefringence (\ref{nk=}) in external
electric $E_{ext}=1e5V/m$ and magnetic $B_{ext}=17T$ fields. The contribution
of the non-local terms\cite{ross}\cite{rothbi} to $\Delta n$, leading to
$\Delta n\propto1/\lambda$, can significantly increase the value of
(\ref{v=int}). However, taking into account that $\Delta n\approx\chi
_{xy}-\chi_{yx}\approx\chi_{xy}+\chi_{yx}$, the $\beta_{\perp}\approx0.1a.u.$,
which corresponds to the experimentally observed $\Delta n\approx
1e-11$\cite{rothbi} according to (\ref{dk=}), is in the range of theoretical
predictions\cite{lyons}\cite{graham}. Therefore $\Delta n$ is assumed to be
constant in the (\ref{v=2}) integral.

According to (\ref{v=int}) $v_{vac}\approx50nm/s$ in external fields
$E_{ext}=1e5V/m$ and $B_{ext}=17T$. The cut-off frequency $\omega_{cut}$ was
chosen to correspond to a wavelength $\lambda=2\pi c/\omega\approx0.1nm$,
since for higher frequencies the molecular polarization vanishes. Density
$\rho\approx1e3kg/m^{3}$, $\Delta n\approx1e-11$ and dielectric constant
$\varepsilon\approx1.5$ were assumed. The contribution of the static field
(\ref{rv=e}) is $v_{stat}\approx20nm/s$. In JRL experiment the estimated
velocity from (\ref{rv=e}) was $v_{laser}\approx1e-15m/s$ (the laser beam
intensity was about $1e5W/m^{2}$). Velocity $v_{vac}$ is not only greater than
$v_{stat}$\ and $v_{laser}$, but may also possess the opposite sign. The
latter follows from substitution of ref. \cite{lyons} results into
(\ref{dk=}). The experimental measurement of (\ref{v=int}) requires effective
homogeneity of the matter. Otherwise eqs. (\ref{rv=e}) and (\ref{rv=}) are not
valid. Thus the region of the crossed fields must be produced locally, similar
to the laser beam in JRL experiment. It can be done for instance by immersing
a capacitor's electrodes inside the liquid.

Any liquid possess viscosity, causing decay in time of any artificially
created flow. The dissipation processes can be taken into account empirically
by adding correspondent viscous forces $\eta\Delta v$ to the equation of
motion (\ref{ddlv}):%
\begin{equation}
\rho\frac{\partial v}{\partial t}=\eta\Delta v \label{rdv}%
\end{equation}
while (\ref{rv=e}) defines the boundary conditions. According to (\ref{rv=e}),
the Gaussian laser beam in JRL experiment generated Gaussian flow. Eq.
(\ref{rdv}), in such case, becomes an ordinary diffusion equation. The
initially gained momentum of the liquid "diffuses" in a plane perpendicular to
the propagation direction. Thus the characteristic decay time is $t_{d}\approx
l_{mirror}^{2}\rho/\eta\approx10s$, where the mirror's size $l_{mirror}%
\approx0.25cm$ and the density to viscosity ratio $\rho/\eta\approx
200s/cm^{2}$\cite{jones2}. Radiation pressure in liquids was studied on the
time scales comparable with $t_{d}$\cite{jones2}. Longer timescale
observations can separate the electromagnetic and liquid flow contributions to
the measured force, because the measured momentum (\ref{p=r}), in the case of
static liquid, corresponds to Abraham's momentum. It is also feasible in an
experiments with strong overlap between incident and reflected beams, e.g.
measurement in a tube. The motion of matter can not occur under symmetric
illumination, thus the measured force must be independent of the liquid's
refractive index.

In conclusion relativistic formalism was applied for light-matter Lagrangian
derivation. Equations of motion were obtained and their correspondence to the
Abraham-Minkowski controversy related experimental data was demonstrated. The
received results correspond to Abraham's predictions, while Minkowski's
momentum can be obtained from (\ref{l=l}) without its last "motion of the
matter" term. Therefore the origin of the controversy lies in the
underestimation of the fact that the field-matter interaction is impossible
without the motion of the latter. The vacuum fluctuations induced flow in
dielectric liquids with $v_{vac}\approx50nm/s$ was predicted in external
crossed electric and magnetic fields. The significant property of this
phenomenon is the high frequency vacuum modes contribution, similar to the
Lamb shift effect. It can be used in future as an investigating tool for zero
fluctuations. Other possible applications lie in fields of microfluidics or
precise positioning of microobjects, e.g. cold atoms or molecules. Initial
experimental verifications can be based on artificially created random fields.

The helpful discussions with A. Englander, B. Sfez and Y.
Siberberg are gratefully acknowledged.


\begin{thebibliography}{99}
\bibitem {smol}In the case of small particles or even molecules the
conservation law (\ref{p=r}) coincides with (\ref{rv=e}), where $E\left(
\varepsilon\mu-1\right)  /4\pi\mu c$ is replaced by the correspondent
polarization. It can be easily shown for the small spherical or cylindrical
particles in the external fields. Although the assumption about matter
homogeneity is not valid and (\ref{ddlv}) can not be used, intuitively it can
be understood considering the Lorentz force acting on a small object in
alternating fields.
\end{thebibliography}
\end{document}